\documentclass[epj]{svjour}
\usepackage{graphics}
\usepackage{graphicx}
\usepackage{color}
\usepackage{bm}
\begin{document}

\title{Magnetic and quantum entanglement properties of the distorted diamond chain model for azurite}
\author{Nerses Ananikian\inst{1,2} \and Hrachya Lazaryan\inst{1} \and Mikayel Nalbandyan\inst{1}
}                     
\offprints{}          
\institute{A.I. Alikhanyan National Science Laboratory, Alikhanian Br. 2, 0036 Yerevan, Armenia \and
Departamento de Ciencias Exatas, Universidade Federal de Lavras, CP 3037, 37200-000, Lavras- MG, Brazil}

\date{Received: date / Revised version: date}
%
\abstract{
We present the results of magnetic properties and entanglement of the distorted diamond chain model for azurite using pure quantum exchange interactions. The magnetic properties and concurrence as a measure of pairwise thermal entanglement have been studied by means of variational mean-field like treatment based on Gibbs-Bogoliubov inequality. Such a system can be considered as an approximation of the natural material azurite, Cu$_3$(CO$_3$)$_2$(OH)$_2$. For values of exchange parameters, which are taken from experimental results, we study the thermodynamic properties, such as azurite specific heat and magnetic susceptibility. We also have studied the thermal entanglement properties and magnetization plateau of the distorted diamond chain model for azurite.} 
\authorrunning{N. Ananikian \and H. Lazaryan \and M. Nalbandyan }
\titlerunning{Magnetic and quantum entanglement properties of azurite }
\maketitle
\section{Introduction}
\label{intro}
Copper oxide materials as low-dimensional magnetic systems are interesting subjects to investigate because of the new physics that can arise in low temperatures.    The natural material azurite Cu$_3$(CO$_3$)$_2$(OH)$_2$ is the one of such copper oxide material
which has been the subject of debates in recent years \cite{1,ricter,richter1,kikuchi,gusu,rule,5}.

\begin{figure}[t]
\includegraphics[width=250pt]{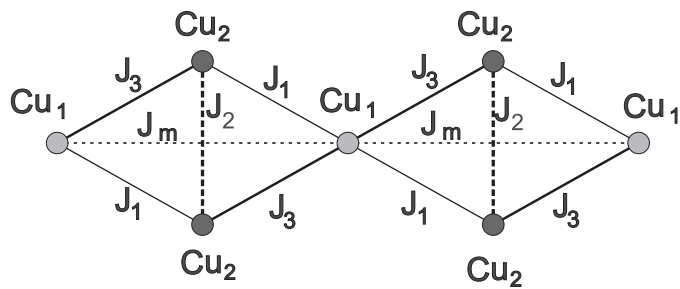}
\caption{\label{diamond1}Distorted diamond chain.}
\end{figure}
Moreover, azurite can be considered as one of the
first experimental realizations of the 1D distorted ($J_1\neq J_3$) diamond
chain model (see Figure~\ref{diamond1}, for the detailed structure of azurite see  for example \cite{ricter}). It shows antiferromagnetic behavior at temperatures below Neel
temperature 1.9 K \cite{2}. Both  the experimental analysis and theoretical modeling have been performed, but there is no clarity about exact values of the relative exchange interactions between azurite's copper ions  \cite{1,ricter,richter1,kikuchi,gusu,rule,5}.   On the other hand, there is significant evidence that the magnitude of the exchange interaction $J_2$ is the greatest one, forming dimers and monomers of Cu- ions \cite{ricter,kikuchi,rule}.
 In \cite{ricter}   an effective generalized spin-1/2 diamond chain model has been suggested with a dominant antiferromagnetic Cu-ions of the dimer coupling $J_2$,  two antiferromagnetic Cu-ions of the dimer-monomer couplings $J_1$ and $J_3$, as well as a significant direct monomer- monomer of Cu-ions  $J_m$  exchange which explains a broad range of experiments on azurite and resolves the existing controversies.

The most noticeable feature is that there is a large magnetization plateau at 1/3 of the saturation magnetization \cite{kikuchi}, which extends from 11 to 30 $ T $, when the magnetic field is applied in the perpendicular to the chain axis. Such 1/3 plateau is usually associated with classical up-up-down $(uud)$  type of spin arrangement (or with a quantum state which has a classical analogue), while the 1/3 plateau of  azurite is proposed to be of fundamentally different, $00u$ type, where the dominant
$J_2$ coupling ensures that the two "dimer" spins on the Cu$_2$ sites (see Figure \ref{diamond1}) are in a singlet state, while the third "monomer" (Cu$_1$) spin is completely polarized by the magnetic field. This state is based on the presence of a singlet and is a pure quantum nature without a classical analogue. Therefore azurite is the first successful candidate for describing a 1/3 quantum plateau state   \cite{Okamoto,gusu2007}.
Another characteristic feature is that there is almost a direct transition from the plateau of saturation, which may be an interpretation as a remnant of the so-called localized magnon-members present in a perfect chain of diamond  \cite{11,12}.

The phenomenon of magnetization plateau has been studied during the past
decade both experimentally and theoretically. The plateau may exist in the magnetization curves of quantum spin systems in the case of a strong magnetic external field at low temperatures. Magnetization plateaus appear in a wide range of models on
chains, ladders, hierarchical lattices and theoretically analysed by dynamical, transfer
matrix approaches as well as by the exact diagonalization in clusters \cite{platoess,platoess1,platoes7,platoes8,platoes88,platoekagome}.
In this paper, we obtain the magnetization plateau in 1D diamond chain using variational mean-field like treatment, based on Gibbs-Bogoliubov
inequality \cite{bogoljub,bogolubov,bogolubov1,bogolubov11,bogolubov12,levon,izvestia}.

 Entanglement \cite{entanglement0,ef2} has gained renewed interest
with the development of quantum information science.  Entangled
states constitute a valuable resource in quantum information
processing \cite{T1,T2} for example the predicted capabilities of
quantum computers rely on entanglement. Numerous different methods of
entanglement measuring have been proposed for its quantification \cite{entanglement0}. In this paper we use concurrence \cite{wooters,wooters1} as entanglement measure of the spin-1/2  system.
Geometrical frustration and thermal entanglement (concurrence) of spin-1/2 Ising-Heisenberg model on a symmetrical diamond chain was studied  in \cite{29}.  In this paper we study the concurrence  of spin-1/2 Heisenberg model on a distorted diamond chain  as the approximation of natural mineral azurite.

This paper is organized in the following way. In section 2 the variational mean-field
formalism based on the Gibbs--Bogoliubov inequality is applied to the Heisenberg
model on the distorted
diamond chain. In section 3 we investigate the magnetic properties of the
system and compare the obtained results with the experimental data  of the magnetization, specific heat and magnetic susceptibility.
In section 4 the concurrence of the system is calculated. Some conclusive remarks are given in the last section.

\section{Gibbs-Bogoliubov approach}

We use  a spin--1/2 Heisenberg
model. The  Hamiltonian of the Heisenberg model  is
\begin{equation}
H = \sum_{\left\langle i,j \right\rangle}J_{i,j}  \vec {S}_i
\vec{S} _j -\sum_ig\mu_B \vec{B}\vec{S} _i,\vspace{-5pt}
\end{equation}
where $\vec{S} _i $
 are the spin--1/2 operators, $J_{i,j}$ is the exchange interaction constants
connecting sites $i$ and $j$,  $\vec B$ is the value of the external
magnetic field, $g$ - the gyromagnetic
ratio   and $\mu_B$ - the Bohr magneton. The Hamiltonian for the distorted diamond chain  can be written as
\begin{equation}
H =J_1\sum\limits_{i} \left[
\bm\alpha_i-\frac{h}{2}\left(S^i_1{}^z+S^i_4{}^z\right)-h\left(S^i
_2{}^z+S^i_3{}^z\right)\right],\label{Hkagome}
\end{equation}
where  $h=g\mu_B B^z$ and the  $g$ is set to 2,06 \cite{ohta}
 and  \begin{eqnarray}
  \bm\alpha_i&=&\vec{ S}^i _2 \vec{S}^i _3  +\delta_m \vec{ S}^i _1 \vec{S}^i _4  +\delta_2(\vec{S}^i _1 \vec{S}^i _3+\vec{S}^i _2 \vec{S}^i _4 )+ \nonumber\\
&+&\delta_3 (\vec{S}^i _1 \vec{S}^i _2+\vec{S}^i _3 \vec{S}^i
_4).\label{alphabeta}\vspace{-4pt}
\end{eqnarray}
where $\delta_2=J_1/J_2,\ \delta_3=J_3/J_2$ and $\delta_m=J_m/J_2$. Here and further  exchange parameters $(J_1,J_2, J_3)$ and the magnetic field $h$ are taken in
Boltzmann' constant scaling i.e. Boltzmann constant is set to be $k_B = 1$.

 Gibbs-Bogoliubov
inequality \cite{bogoljub,bogolubov,bogolubov1,bogolubov11,bogolubov12,levon,izvestia}
 states that for  free energy (Helm\-holtz
potential) of the system  we have
\begin{equation}\label{bogolubovinequality}
F \le F_0  + \left\langle {H - H_0 } \right\rangle _0,
\end{equation}
where $H$ is the real Hamiltonian which describes the system and $H_0$ is the trial one. $F$
and $F_0$ are free energies corresponding to $H$ and $H_0$ respectively and $ \left\langle \dots  \right\rangle _0$ denotes the
thermal average over the ensemble defined by $H_0$.

By introducing trial
Hamiltonian for our model  containing unknown  variational parameters one can  minimize right hand side of Gibbs-Bogoliubov inequality (\ref{bogolubovinequality}) and get the values of those parameters.

\begin{figure}[t]
\includegraphics[width=255pt]{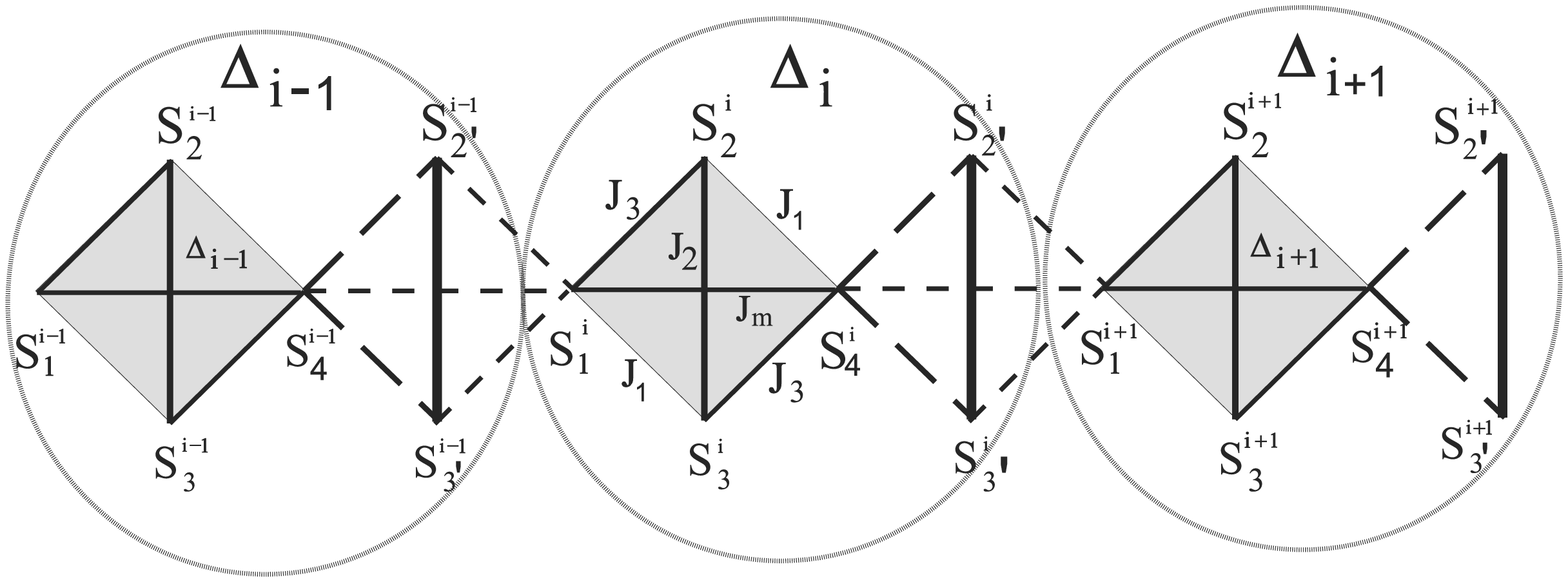}
\caption{\label{diamond2} Each $\Delta_i$ cluster consists of one rectangle
of $\vec{ S}^i _1$, $ \vec{S}^i _2$, $ \vec{S}^i _3$, $ \vec{S}^i _4$ sites
(grey rectangle) and
dimer of $\vec{ S}^i _{2'}$, $ \vec{S}^i _{3'}$ sites (bold line). }
\end{figure}

We introduce a trial Hamiltonian $ H_0 $ as a set of noninteracting
clusters of two types  in the  external self-consistent field: the rectangle and the line (see Figure \ref{diamond2})
\begin{equation}\label{trialsum} H_0  = \sum\limits_{\Delta_i}
{(H_0{}^{(i)}  +H'_0{}^{(i)}  }),
\end{equation}
where the indexes of summation
  $\Delta_i$ label different noninteracting clusters and
\begin{eqnarray}
H_0^{(i)} &=& \lambda \left(\bm \alpha_i \right) - \gamma_1 S^{i}_1{}^z-
\gamma_{2}S^{i}_2{}^z-
\gamma_3S^{i}_3{}^z-\gamma_4S^{i}_4{}^z,\nonumber\\
H'_0{}^{(i)}&=&\lambda' \left(\vec{ S}^i _{2'} \vec{S}^i _{3'}
\right)- \gamma'_{2}S^{i}_{2'}{}^z-
\gamma'_3S^{i}_{3'}{}^z,\label{Htrial}
\end{eqnarray}
where $\lambda,\lambda'$ and $\gamma_j,\gamma'_j$ are the variational parameters.

The real Hamiltonian (\ref{Hkagome})
 can be represented in the following form:
\begin{equation}\label{sum}
H  = \sum\limits_{\Delta_i} {H^{(i)}  },
\end{equation}
where $H^{(i)}$ has the following form:
\begin{eqnarray}
H^{(i)}&=& J_2\bm\alpha_{i}-{h}\left(S^i _1{}^z+S^i _2{}^z+S^i_3{}^z+S^i_4{}^z\right)+J_2\left(\vec{ S}^i _{2'} \vec{S}^i _{3'}  \right)-\nonumber\\
&-&h(S^i _{2'}{}^z+S^i_{3'}{}^z)+\frac{J_m}{2}(\vec{S}^i _1
\vec{S}^{i-1} _4 + \vec{S}^i _4 \vec{S}^{i+1}
_1)+\nonumber\\&+&J_1(\vec{S}^i _1 \vec{S}^{i-1} _{3'}+ \vec{S}^i _4
\vec{S}^{i} _{2'})+J_3(\vec{S}^i _1 \vec{S}^{i-1} _{2'}+ \vec{S}^i
_4 \vec{S}^{i} _{3'}),
\end{eqnarray}
Gibbs-Bogoliubov inequality (\ref{bogolubovinequality}) can be rewritten now as follows:
\begin{equation}\label{inequality}\begin{array}{l}\vspace{5pt} f^{(i)}  \le f^{(i)}_0
+\left(J_2-\lambda\right)
\left\langle\bm\alpha_{i}\right\rangle_0
+(J_2-\lambda ')\left\langle\vec{ S}^i _{2'} \vec{S}^i _{3'}  \right\rangle_0+\\
+\ J_{1}(m_1 m'_3+m_{4}m'_{2})+J_3(m_1 m'_2+m_4m'_3)+\\+\ J_{m}(m_1m_4)-\ \sum\limits_{j=1}^4\left({h}-\gamma_j\right)m_j-\sum\limits_{j=2}^3(h-\gamma'_j)m'_j,
\end{array}\end{equation}
where
 $f^{(i)}=F/N$ and $f_0^{(i)}=F_0/N $ are free energies of the one  cluster
 ($N$ is number of clusters) and
 we denote by $m_j\equiv\left\langle S^i _j{}^z \right\rangle_0$ and
 $m'_{j'}\equiv\left\langle S^i _{j'}{}^z \right\rangle_0$ the  magnetisations
of $\Delta_i$ cluster.
Here we take into account that spins of different clusters are statistically independent, i.e. $\left\langle\bm{S}^i_j \bm{S}^{i,i\pm1}_{k'}\right\rangle_0=m_jm'_{k}$.
For given $h,J_i$ one must perform the minimization of r.h.s
and obtain the values for $\gamma,\lambda$ and $\gamma',\lambda'$. Minimizing the right hand side  of (\ref{inequality}) in order to $\gamma_j,\lambda$ and $\gamma'_j,\lambda'$
and using the fact, that $\displaystyle\frac{{\partial f^{(i)}_0}}{{\partial \lambda}}=\left\langle\bm\alpha_i\right\rangle_0$
and $\displaystyle\frac{{\partial f^{(i)}_0}}{{\partial \gamma_j}}=-m_j$ we obtain the following values for
the variational parameters:
\begin{eqnarray}
\lambda&=&\lambda'=J_2,\nonumber\\
\gamma_2&=&\gamma_3=h,\nonumber\\
\gamma_1&=&h-J_1m'_3-J_3m'_2-J_mm_4,\nonumber\\
\gamma_4&=&h-J_3m'_3-J_1m'_2-J_mm_1,\nonumber\\
\gamma'_2&=&h-J_1m_4-J_3m_1,\nonumber\\
\gamma'_3&=&h-J_3m_4-J_1m_1.\label{variablse}
\end{eqnarray}
Using this values and the trial
Hamiltonian one can calculate the value of any thermodynamical function of
our system for the fixed $h,J_i$.

\section{Magnetisation, specific heat and susceptibility}
\begin{figure}[t]
 \includegraphics[width=240pt]{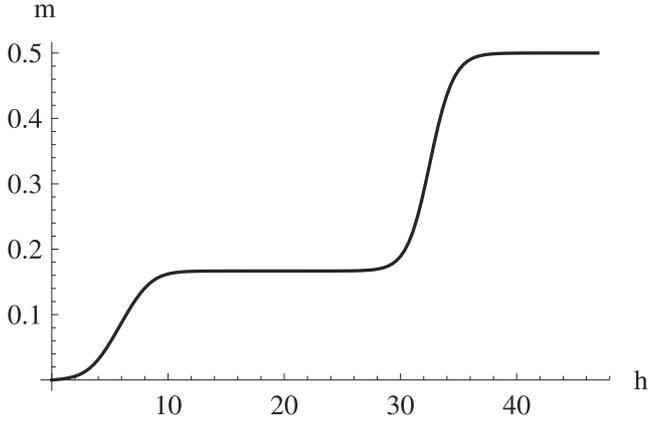}
\caption{ \label{magnet1} Magnetization $m$  of the cluster  versus external magnetic field $h$ (Tesla) for
$J_2=33K,$ $J_1 = 15.51\ K$, $J_3=6.93\ K, J_{m}= 4.62\
K$ at  and  T=1.3 K.   }
\end{figure}

The results of the previous section can be used for investigation of the magnetic
properties of the model. The magnetization of arbitrary site $m_\upsilon$
of cluster $\Delta_i$ is defined as
\begin{equation}\label{magndef}
m_\upsilon=\frac{Tr( S^i_\upsilon{}^z e^{ -H_0^{(i)}/T})}{Z}
\end{equation}
where $S^i_\upsilon{}^z$ is corresponding spin operator, $H_0^{(i)}$
is the Hamiltonian (\ref{Htrial}) and $Z$ is the corresponding
partition function of the cluster. To obtain all six magnetizations
($ m'_2,m'_3$, $m_1,\dots m_4$) one must insert  the values of
the variation parameters  (\ref{variablse}) into the equation
(\ref{magndef}) and solve  the resulting system of equations for the
fixed $h,J_i$.
\begin{equation}\label{set12}
\left\{
\begin{array}{rcl}
m_1&=&f_1(m'_2,m'_3,m_1,m_4)\\
\dots\\
m_4&=&f_4(m'_2,m'_3,m_1,m_4)\\
m'_2&=&f'_2(m_1,m_4)\\
m'_3&=&f'_3(m_1,m_4)\\
\end{array}
\right.
\end{equation}
 The calculations of our paper is based on the effective diamond
chain with  values of coupling constants taken from  \cite{ricter}:
\begin{equation} J_2=33K;
\ \delta_2=15.51/33;\ \delta_3=6.93/33;\ \delta_m=4.62/33.
\end{equation}

The dependence of the average magnetization (for cluster)
\begin{equation}\label{average}
m=\frac{m_1+m_2+m_3+m_4+m'_2+m'_3}{6}
\end {equation}
from external magnetic field,  calculated
from (\ref{set12}), is shown
in Figure \ref{magnet1}. As it can be seen from the Figure~\ref{magnet1}  the 1/3 magnetisation plateau at $T=1.3K$ extends from
11 $T$ to 29 $T$  interval, while the  experimental   data \cite{ricter,kikuchi}
gave $11\ T-30\ T$ interval.

\begin{figure}
 \includegraphics[width=240pt]{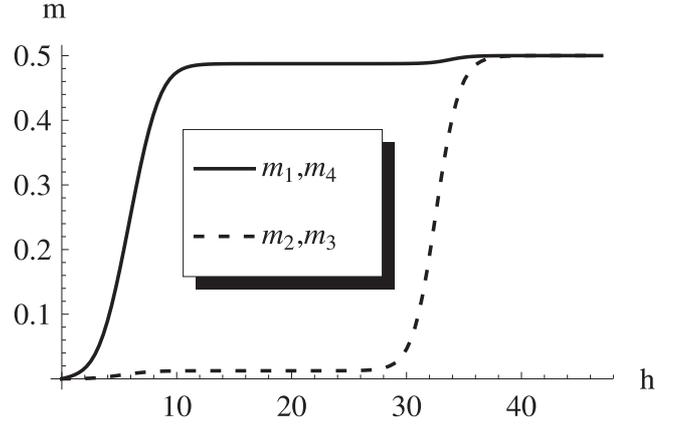}
 \caption{\label{magnet2}
Magnetizations $m_1,m_2,m_3$ and $m_4$  versus external magnetic field $h$ (Tesla) for $J_2=33K,$ $J_1 = 15.51\ K, J_3=6.93\ K, J_{m}= 4.6\
K.$ at  T=0.8 K.}
\end{figure}

As it mentioned above, azurite is a good candidate to exhibit  $00u$ type  1/3 plateau state \cite{Okamoto,gusu2007}.
The plots in Figure~\ref{magnet2}  illustrate magnetisation
versus  magnetic field dependencies    for different sites obtained using
mean-field approach.
 As it can be seen from the Figure \ref{magnet2} dimers are essentially in the singlet
state (dashed lines) whereas the single "monomer" spins are almost
fully polarized in the 1/3 plateau (solid lines).  The observed
polarization is incompatible with a $uud$ type of
plateau, in which the dimer spins are strongly polarized. We find
that dimer spins are about 2.5\% polarized  while the numerical calculation
for the  ideal diamond chain Heisenberg model gives 2.7\% polarisation \cite{ricter},
while experiments give 10\% \cite{Okamoto}.
\begin{figure}
 \includegraphics[width=240pt]{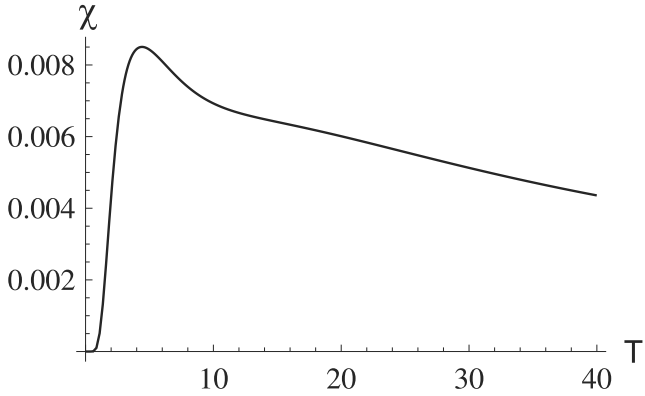}
 \caption{\label{sescept1}
Zero field magnetic susceptibility versus temperature.  }
\end{figure}

The magnetic susceptibility
is defined as follows:\begin{equation}
\chi_0=\left(\frac{\partial m}{\partial h}\right)_{h=0}.
\end{equation} The  magnetic susceptibility    measurements of azurite performed in \cite{kikuchi} and
it was found double-peak-like structure in the magnetic susceptibility  curve,
namely, in the temperature dependence of the magnetic
susceptibilities  the sharp peak appears at $5 K$ and the rounded peak is observed at
$23 K$. The  Figure \ref{sescept1} shows the temperature
dependence of magnetic susceptibility obtained by our calculations using
(\ref{set12}) and (\ref{average}). It has first sharp peak at  $4.4K$.

\begin{figure}
 \includegraphics[width=240pt]{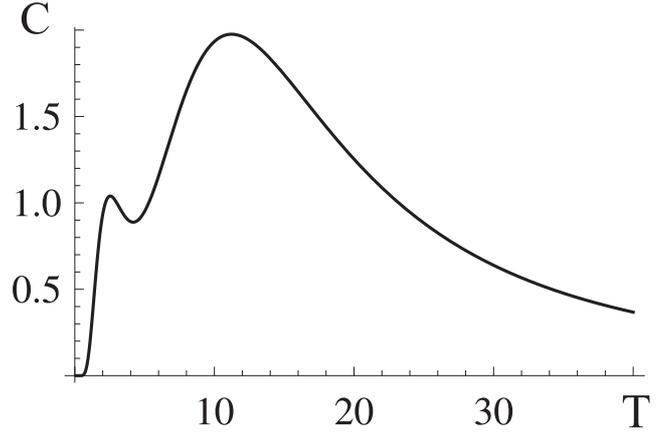}
 \caption{\label{specific1}
Zero field specific heat versus temperature.  }
\end{figure}

Analogous to the magnetic susceptibility, we also have
calculated specific heat:\begin{equation}
C=-T\left(\frac{\partial ^{2}f_{0}}{\partial T^2 }\right)_{h=0}.
\end{equation}
The specific heat measurements  for azurite
are performed in  \cite{kikuchi} and a sharp peak
is observed at $T_N=1.8 K$  signaling
an ordering transition  and two anomalies have been observed
in the specific heat at $T = 4 K$  and $T = 18 K$ \cite{kikuchi,kkk}.
The first peak  is out of reach of a one-dimensional
model.    Our calculations gave a low-temperature
peak  for $H = 0$ at $T=3$K and the second peak at $12 K$ (Figure
\ref{specific1}). The obtained double-peak-like structure of specific heat   reproduce the important features of the experimental results \cite{ricter,kikuchi}.  Density functional theory (DFT) \cite{ricter} also gives the double-peak-like structure for the specific heat.   In this paper we have reproduced the important features of the experimental results in the specific heat properties of azurite theoretically.
\section{Concurrence and thermal entanglement}
 The mean-field like treatment  transforms many-
body system to the set of clusters in the effective self-consistent magnetic
field. This allows to study, in particular, thermal entanglement properties
of the system. The entanglement of formation  \cite{ef1,ef2} was proposed to quantify the entanglement
of a bipartite system. Unfortunately  the entanglement of formation  extremely difficult to calculate,
 in general.
However, in the special case of two spin 1/2 particles an analytical expression    \cite{wooters,wooters1}
can be obtained for the entanglement of formation   of any density matrix of two
spin 1/2 particles: \begin{equation}
EF=H(\frac{(1+\sqrt{1-C^{2}})}{2}),
\end{equation}
where
\begin{equation}
H(x)=-x\log_2(x)-(1-x)\log_2(1-x),
\end{equation}
and $C$ is the quantity called \textit{concurrence} \cite{wooters,wooters1}.
 In the case of diamond chain  the concurrence
$C_{i,j}$ corresponding to the pair $(i,j)$ has the following form:
\begin{equation}\label{18}\nonumber
C_{i,j}=max\{\lambda_1-\lambda_2-\lambda_3-\lambda_4,0\},
\end{equation}
where $\lambda_k$ are the square roots of the eigenvalues ($\lambda_1$ is
the maximal one) of the
matrix
\begin{equation}\label{19}\nonumber
\tilde\rho=\rho_{ij}\cdot(\sigma_i^y\otimes\sigma_j^y)\cdot\rho^*_{ij}\cdot(\sigma_i^y\otimes\sigma_j^y),
\end{equation}
where $\rho_{ij}$ is the
reduced density matrix for  $(i,j)$ pair. For rectangle cluster the density matrix is
\begin{equation}
\rho=\frac{1}{Z}\sum\limits_{i=1}^{16}e^{-\frac{E_i}{T}}|\psi_{i}\rangle\langle\psi_{i}|,
\end{equation}where $Z$ is the partition function of the rectangle  and $\psi_i$ and $E_i$
are eigenvectors and eigenvalues of the Hamiltonian  $H_0^{(i)}$
(\ref{Htrial}) respectively. The reduced density matrix of  $(i,j)$ pair     $\rho_{ij}$ can be calculated
as\begin{equation}
\rho_{ij}=\sum_{p}\left\langle\phi^{kl}_p|\rho|\phi^{kl}_p\right\rangle,
\end{equation}   where $\phi^{kl}_p$ is the $p$-th basis vector for $(k,l)
$
remaining pair of sites of rectangle. After the calculations it has the following form
\begin{equation}\label{21}\nonumber
\rho_{ij}=
\left(\begin{array}{cccc}
u&0&0&0\\
0&w&y&0\\
0&y&w&0\\
0&0&0&v
\end{array}\right),
\end{equation}where $u,w,y$ and $v$ are some functions of variables $\gamma,\lambda,\gamma',\lambda'$ and $T$. Using (\ref{18}), (\ref{19}) and (\ref{21})  one can find the following expression for  the concurrence $C_{i,j}$
\begin{equation}\label{concur}
C_{i,j}=2 max\{|y|-\sqrt{uv},0\}.\nonumber
\end{equation}
 \begin{figure}[t]
$\begin{array}{cc}
 \includegraphics[width=230pt]{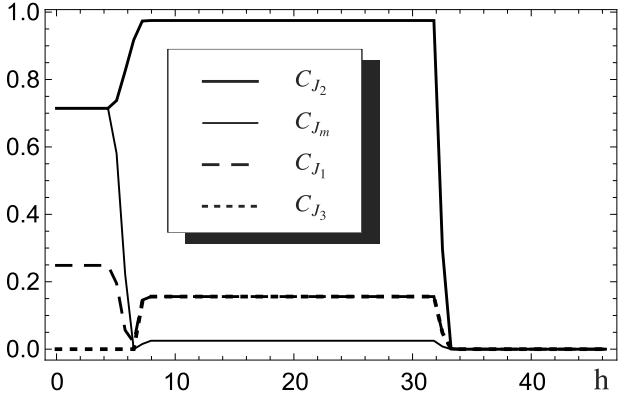}\\
\end{array}$\caption
{   The concurrences $C_{J_2},C_{J_1},C_{J_3}$ and $C_{J_m}$ versus external magnetic field $h$ (Tesla) at $T=0.1K$  for $J_2=33K,$ $J_1 = 15.51\ K, J_3=6.93\ K, J_{m}= 4.62\
K.$ \label{figconcur} }
\end{figure}In Figure \ref{figconcur} is shown the dependence of concurrences  $C_{2,3}\equiv C_{J_2},\ C_{1,2}= C_{3,4}\equiv C_{J_3},\ C_{1,3}= C_{2,4}\equiv C_{J_1}$ and $C_{1,4}\equiv C_{J_m}$ on
the magnetic field.  The behavior of the concurrence can be used to analyze spin phases of azurite.  As it can be seen from the Figure  \ref{figconcur} there is  three regions in magnetic field axes with different ground states.
 For lower value of external magnetic field the opposite spins ($S_2$ and $S_3$) in diamond cluster is highly entangled. The neighboring spins with lower coupling constant are not entangled.
For higher values of $h$ when the magnetization has a plateau the entanglement
of  ($S_1,S_4$) pair is almost zero, i. e. practically unentangled, while the ($S_2$, $S_3$) pair is almost  fully entangled. The concurrence of the neighboring spins  on the plateau  is  small, comparing to ($S_2$, $S_3$) pair, moreover
the neighboring spins with lower coupling constant ($J_3$)  are unentangled.
 \begin{figure}[t]
$\begin{array}{cc}
 \includegraphics[width=230pt]{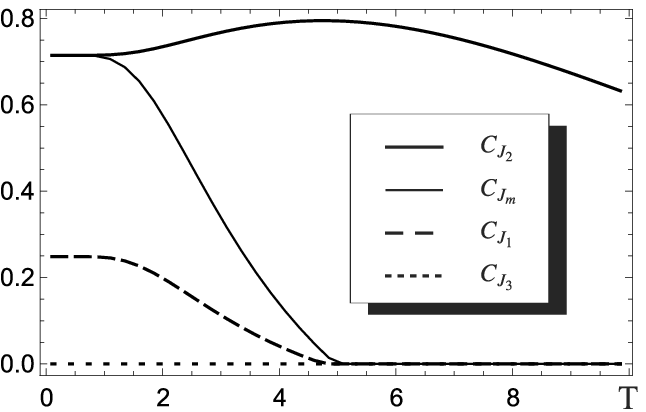}\\
 a)\\
 \includegraphics[width=230pt]{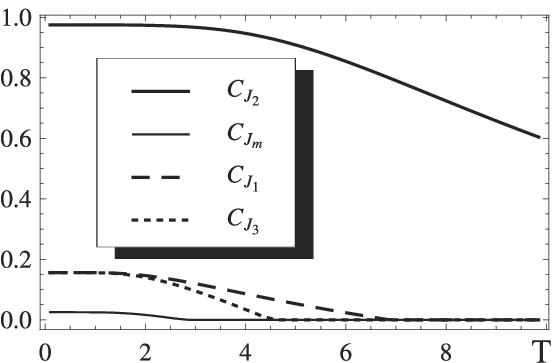}\\
 b)
\end{array}$\caption
{   The concurrences $C_{J_2},C_{J_1},C_{J_3}$ and $C_{J_m}$ versus temperature for a) $h=0T$ and b) $h=18T$. \label{figconcurT} }
\end{figure}

Now, we consider the dependence of the concurrence on
temperature. Figure \ref{figconcurT} (a) shows the temperature dependence of the concurrence for
different pairs of diamond chain at  small value of the magnetic field ($h=1T$). The neighboring spins with lower coupling constant ($J_3$) stay unentangled with increasing temperature while the concurrence of the bigger one ($J_1$) decreases with temperature to $4.5K,$ where the
entanglement vanishes. The concurrences of  $J_m$ pair  $J_1$
behave similar  and vanish almost at the same temperature, while the dominant $J_2$ pair stays entangled at higher temperatures (until  $28K$).  Almost the
same behaviour shows the temperature dependence of the concurrence for
different pairs of diamond chain at plateau phase $h=18T$ (Figure \ref{figconcurT} (b)). The concurrences
for $J_m,J_{3}$ and $J_1$  decrease with temperature and vanish sequential
between $4K$ and $7K$ while
the concurrence
for $J_2$ pair stays entangled for the higher  temperatures and vanishes
at the same temperature as for small values of magnetic field  ($28K$). \\
As it can be seen from the Figure \ref{figconcur} and \ref{figconcurT}b in the plateau state dimers are almost fully entangled (lines labeled by $C_{J_2}$ in the figures) whereas the monomer spins are weakly entangled (lines labeled by $C_{J_1}$ and $C_{J_3}$
in the figures).  The observed
entanglements is incompatible with a ($uud$) type of
plateau, and confirm the proposed ($00u$) nature of the plateau.

Now we revert to the Figure \ref{figconcurT} (a) to notice that
comparison of the Figure \ref{figconcurT} (a) with the Figure
\ref{specific1} shows that the $C(J_2)$ has a peak at nearly
$T=5K$ and it is located between two peaks of the specific heat.
Roughly such a behavior can be understood as follows.\\
As a result of interaction between the horizontal ($J_m$) and
vertical ($J_2$) dimers and also as a result of an asymmetry
$(J_1>J_3)$, decreasing of the concurrences (in comparison with
non-interacting case) of these dimers at zero temperature is
observed. As temperature increases, energy "accumulates" in the
horizontal dimer at first and in the vertical dimer later, which
causes the double peak structure in the specific heat picture (see
Figure \ref{specific1}). During the process of "energy
accumulation" in the sites of the horizontal dimer (the first peak
region) destruction of a quantum correlations between them takes
place as a result of thermal fluctuations (so the concurrence
$C(J_m)$ is decreasing, see Figure \ref{figconcurT} (a)). And also
destruction of a quantum correlations between the sites of
horizontal dimer and vertical one occurs. As a consequence one can
see an increasing of $C(J_2)$ from $T=0K$ to $T=5K$ (without above
mentioned couplings (non-interacting dimers case) $C(J_2)$ gets
its maximal value equals to $1$ in this region). Further
temperature increasing brings destruction of the quantum
correlations between the sites of the vertical dimer, which is the
result of "energy accumulation" on these sites (and as a
consequence - decreasing of $C(J_2)$ and increasing the
specific heat to the second peak).\\
Varying values $J_1, J_2, J_3, J_m$ of the coupling constants,
brings to analogical picture, except the symmetric case where $J_1
= J_3$. In the symmetric case the concurrences $C(J_2)=C(J_m)=1$
(at zero temperature) and are decreasing with the temperature and
as fast as are higher the values of the $J_1=J_3$ coupling
constants. Closer is the diamond to the symmetric case, worse is
the appearance of the peak structure in the $C(J_2)$ picture.

  \section{Conclusions}
In this paper using Heisenberg model the distorted diamond chain
was studied as approximation for natural material, azurite
Cu$_3$(CO$_3$)$_2$(OH)$_2$.  The magnetic properties and
concurrence as a measure of pairwise thermal entanglement of the
system was studied by means of variational mean-field like
treatment based on Gibbs-Bogoliubov inequality.  In our approach
for the values of exchange parameters taken from theoretical and
experimental results we have obtained the 1/3 magnetization
plateau with correct critical values of magnetic field, moreover
this plateau is caused   by   $00u$ type plateau state. We also
studied the thermal entanglement properties of the distorted
diamond chain and drew a parallel between them and the specific
heat ones.

\section*{Acknowledgments}
We would like to thank Prof. Johannes Richter for helpful discussions, comments and valuable suggestions.
This work has been supported by the French-Armenian grant No. CNRS IE-017   and by the Brazilian FAPEMIG grant No. CEX - BPV - 00028-11.


\begin{thebibliography}{30}
%
%
%
\bibitem{1} A. Honecker,\  A. Lauchli, Phys. Rev. B \textbf{63},
174407 (2001)
\bibitem{ricter}H. Jeschke, I. Opahle, H. Kandpal, R. Valenti, H. Das, T.~Saha-Dasgupta, O. Janson, H.~Rosner, A.~Bruhl, B.~Wolf, M.~Lang, J.~Richter, S.~Hu, X.~Wang, R.~Peters, T.~Pruschke, A.~Honecker, Phys. Rev. Lett. \textbf{106},
217201 (2011)
\bibitem{richter1}A. Honecker, S. Hu, R. Peters, J. Ritcher, J. Phys.: Condens. Matter\textbf{ 23,}  164211 (2011)\bibitem{kikuchi}
 H. Kikuchi, Y. Fujii, M. Chiba, S. Mitsudo, T. Idehara, T.
Tonegawa, K. Okamoto, T. Sakai, T. Kuwai,  H. Ohta,
Phys. Rev. Lett.\textbf{ 94},  227201 (2005)
\bibitem{kkk}J. Kang, C. Lee, R. K. Kremer, M-H Whangbo, J. Phys.: Condens. Matter \textbf{21,}  392201 (2009)
\bibitem{gusu}
B. Gu,  G. Su, Phys. Rev. Lett. \textbf{97},  089701 (2006)
\bibitem{rule} K. C. Rule, A. U. B. Wolter, S. S¨ullow, D. A. Tennant,
A.~Br¨uhl, S. K¨ohler, B. Wolf, M. Lang,  J. Schreuer,
Phys. Rev. Lett. \textbf{100}, 117202 (2008)
\bibitem{5}
 H.-J. Mikeska, C. Luckmann, Phys. Rev. B \textbf{77},  054405 (2008)
\bibitem{2} R. D. Spence, R.D. Ewing, Phys. Rev. \textbf{112},
1544 (1958)
\bibitem{Okamoto} K. Okamoto, T. Tonegawa, M. Kaburagi, J. Phys.:
Condens. Matter \textbf{15,}  5979 (2003)
\bibitem{gusu2007}
B. Gu, G. Su, Phys. Rev. B \textbf{75},  174437 (2007)
\bibitem{11}
J. Schulenburg, A. Honecker, J. Schnack, J. Richter,  H.-J. Schmidt, Phys. Rev. Lett. {\bf 88},   167207 (2002)
\bibitem{12} O. Derzhko and J. Richter, Eur. Phys. J. B {\bf 52},
23  (2006)
\bibitem{platoess}T. A. Arakelyan, V. R. Ohanyan, L. N. Ananikian, N.~S.~Ananikian, M. Roger,
Phys. Rev. B \textbf{67},  024424 (2003)
\bibitem{platoess1}G. Japaridze,  S. Mahdavifar, Eur. Phys. J. B \textbf{68},  59 (2009)
\bibitem{platoes7}V.V. Hovhannisyan, L.N. Ananikyan,   N. S.  Ananikian, Int. J. of Mod. Phys. B  \textbf{21},  3567 (2007)
\bibitem{platoes8} V.V. Hovhannisyan, N.S. Ananikian,  Phys. Lett. A \textbf{372},
 3363 (2008)
\bibitem{platoes88}V. R. Ohanyan, N. S. Ananikian, Phys. Lett. A \textbf{307}  76 (2003)
\bibitem{platoekagome}N. Ananikian, L. Ananikyan, R. Artuso, H. Lazaryan, Phys. Lett. A \textbf{374,}   4084 (2010)
\bibitem{bogoljub} N. N. Bogoliubov  J. Phys. (USSR) \textbf{11,}  23 (1947)
\bibitem{bogolubov} G. D. Mahan   {\it Many-Particle Physics}
(New York: Kluwer/Plenum 2000)
\bibitem{bogolubov1} S-S Gong, G. Su   Phys. Rev. A \textbf{80,}  012323
(2009)
\bibitem{bogolubov11} M. Asoudeh,  V. Karimipour    Phys. Rev. A. \textbf{73,}   062109 (2006)
\bibitem{bogolubov12} N. Canosa, J. M.   Matera, R. Rossignoli,
Phys. Rev. A. \textbf{76,}  022310 (2007)
\bibitem{levon}N. S. Ananikian, L. N.    Ananikyan, L. A. Chakhmakhchyan,
 A. N. Kocharian  J. Phys. A \textbf{44,}   025001 (2011) \bibitem{izvestia}L. Ananikyan, H. Lazaryan,
 Journal of Contemporary Physics (Armenian Academy of Sciences) \textbf{46,}   184 (2011)

\bibitem{entanglement0}L.  Amico, R. Fazio, A.  Osterloh, V. Vedral    Rev. Mod. Phys. \textbf{80,}  517 (2008) \bibitem{entanglement02}
R. Horodecki, P. Horodecki, M. Horodecki, K.Horodecki,   Rev. Mod. Phys. \textbf{81,}  865 (2009)
\bibitem{ef1}
 C. H. Bennett, D. P. DiVincenzo, J. Smolin,  W. K. Wootters, Phys. Rev. A \textbf{54,}
3824 (1996)
\bibitem{ef2} C. H. Bennett, G. Brassard, S. Popescu, B. Schumacher, J. Smolin,  W. K. Wootters, Phys.
Rev. Lett. \textbf{76,}  722 (1996)
\bibitem{T1}C. H. Bennett, D. P. DiVincenzo, Nature \textbf{404},   247 (2000)
\bibitem{T2}  D. Loss,   D. P. DiVincenzo  Phys. Rev. A \textbf{57,}  120
(1998)
\bibitem{wooters} S. Hill,  W. K. Wootters     Phys. Rev. Lett. \textbf{78,} 5022 (1997)
\bibitem{wooters1}  W. K. Wootters  Phys. Rev. Lett.
\textbf{80,}  2245 (1998)
\bibitem{29}N.  Ananikian, L. Ananikyan, L. Chakhmakhchyan, O. Rojas, (e-print: cond-mat.str-el/1110.6406)
\bibitem{ohta} H. Ohta, et.al,
J. Phys. Soc. Jpn. \textbf{72},   2464 (2003)
\end{thebibliography}
\end{document}